\documentclass{emulateapj}
\usepackage{graphicx}
\usepackage{times}

\def\gsim { \lower .75ex \hbox{$\sim$} \llap{\raise .27ex \hbox{$>$}} }
\def\lsim { \lower .75ex \hbox{$\sim$} \llap{\raise .27ex \hbox{$<$}} }
\def\be{\begin{equation}}
\def\ee{\end{equation}}
\def\bea{\begin{eqnarray}}
\def\eea{\end{eqnarray}}

\newcommand{\plb}{{Phys.~Lett.~B}}

\shorttitle{Testing CPT Symmetry with CMB Measurements: Update
after WMAP5} \shortauthors{Xia {\it et al.}}

\begin{document}

\title{Testing CPT Symmetry with CMB Measurements:
Update after WMAP5}

\author{Jun-Qing Xia\altaffilmark{1}, Hong Li\altaffilmark{2},
Gong-Bo Zhao\altaffilmark{3} \& Xinmin Zhang\altaffilmark{1}}

\altaffiltext{1}{Institute of High Energy Physics, Chinese Academy
of Science, P. O. Box 918-4, Beijing 100049, P. R. China;
xiajq@mail.ihep.ac.cn.}

\altaffiltext{2}{Department of Astronomy, School of Physics,
Peking University, Beijing, 100871, P. R. China.}

\altaffiltext{3}{Department of Physics, Simon Fraser University,
Burnaby, BC, V5A 1S6, Canada.}

\begin{abstract}
In this Letter we update our previous result on the test of CPT
symmetry with Cosmic Microwave Background (CMB) measurements. A
CPT violating interaction in the photon sector ${\cal L}_{cs} \sim
p_{\mu}A_{\nu}\tilde F^{\mu\nu}$ gives rise to a rotation of the
polarization vectors of the propagating CMB photons. Recently the
WMAP group used the newly released polarization data of WMAP5 to
measure this rotation angle $\Delta\alpha$ and obtained
$\Delta\alpha=-1.7\pm2.1$ deg ($1\sigma$). However, in their
analysis the BOOMERanG 2003 data is not included. Here we revisit
this issue by combining the full data of WMAP5 and BOOMERanG 2003
angular power spectra for the measurement of this rotation angle
$\Delta\alpha$ and find that $\Delta\alpha=-2.6\pm1.9$ deg at a
$68\%$ confidence level.

\end{abstract}



\section{Introduction}

The fundamental CPT symmetry which has been proved to be exact in
the framework of the standard model of particle physics and
Einstein gravity could be dynamically violated in the expanding
universe. This type of cosmological CPT violation mechanism
investigated in the literature
\cite{Li:2001st,Li:2002wd,Li:2004hh} has an interesting feature
that the CPT violating effect at present time is small enough to
satisfy the current laboratory experimental limits, but large
enough in the early universe to account for the generation of the
matter anti-matter asymmetry. More importantly, it could be
accumulated to be observable in the cosmological experiments
\cite{Feng:2004mq,Li:2007}. With the accumulation of high quality
data on the CMB measurements, cosmological observations become a
powerful tool to test this fundamental symmetry.

For a phenomenological study in the photon sector the CPT
violation can be parameterized in terms of an effective lagrangian
\cite{Carroll:1989vb,Carroll:1990zs}:
\begin{equation}\label{Eq1}
\mathcal{L} = -\frac{1}{4}F_{\mu\nu}F^{\mu\nu}+\mathcal{L}_{cs}~,
\end{equation}
where $\mathcal{L}_{cs}\sim p_{\mu}A_{\nu}\tilde F^{\mu\nu}$ is a
Chern-Simons term, $p_{\mu}$ is an external vector and $\tilde
F^{\mu\nu}=(1/2)\epsilon^{\mu\nu\rho\sigma}F_{\rho\sigma}$ is the
dual of the electromagnetic tensor. This Lagrangian is not
generally gauge invariant, but the action is gauge independent if
$\partial_{\nu}p_{\mu}=\partial_{\mu}p_{\nu}$. This may be
possible if $p_{\mu}$ is constant in spacetime or the gradient of
a scalar field in the quintessential baryo-/leptogenesis
\cite{Li:2001st,Li:2002wd,quin_baryogenesis} or the gradient of a
function of the Ricci scalar in gravitational baryo-/leptogenesis
\cite{Li:2004hh,R}. The Chern-Simons term violates Lorentz and CPT
symmetries, and also the $P$ and $CP$ symmetries when the
background field $p_0$ does not vanish.

For the CMB measurements the Chern-Simons term induces a rotation
of the polarization \cite{Li:2007,Xia:2007qs} with the rotation
angle $\Delta\alpha$ given by
\begin{equation}\label{Eq2}
\Delta\alpha=\int^{t_0}_{t_{(z\simeq 1100)}} ~ p_0 ~ \frac{dt}{a}~
= p_0~l_{(z\simeq 1100)}~,
\end{equation}
where $p_0$ denotes the time component of $p_{\mu}$ and
$l_{(z\simeq 1100)}$ is the comoving distance of the CMB photon
emitted at the last scattering surface and $t_0$ indicates the
present time. In Eq.(\ref{Eq2}) we have assumed $p_0$ is a
constant. For a more general case please see our previous
companion paper \cite{Xia:2007qs}.

For the standard theory of CMB, the TB and EB cross-correlation
power spectra vanish. In the presence of the CPT violating term
(Eq.(\ref{Eq1})) the polarization vector of each photon is rotated
by an angle $\Delta\alpha$, and one expects to observe nonzero TB
and EB power spectra, even if they are zero at the last scattering
surface. Denoting the rotated quantities with a prime, one gets
\cite{Feng:2004mq,Lue:1998mq}:
\begin{eqnarray}
C_{l}^{'TB} &=& C_{l}^{TE}\sin(2\Delta\alpha)~, \nonumber\\
C_{l}^{'EB} &=&
\frac{1}{2}(C_{l}^{EE}-C_{l}^{BB})\sin(4\Delta\alpha)~,\nonumber\\
C_{l}^{'TE} &=& C_{l}^{TE}\cos(2\Delta\alpha)~,\label{modify}\\
C_{l}^{'EE} &=& C_{l}^{EE}\cos^2(2\Delta\alpha) +
C_{l}^{BB}\sin^2(2\Delta\alpha)~,\nonumber\\
C_{l}^{'BB} &=& C_{l}^{BB}\cos^2(2\Delta\alpha) +
C_{l}^{EE}\sin^2(2\Delta\alpha)~,\nonumber
\end{eqnarray}
while the CMB temperature power spectrum remains unchanged.

In Xia et al. (2007), using the full data of BOOMERanG 2003 and
the WMAP3 angular power spectra we have performed the analysis on
the determination of the rotation angle $\Delta \alpha$ and find
that $\Delta\alpha=-6.2\pm3.8$ deg ($1\sigma$). This result
improves the measurement given by our previous paper
\cite{Feng:2006dp} and the paper by Cabella et al.
(2007)\footnote{For the implications of this measurement on the
possible new physics, please also see papers
\cite{LiuCPT,KosteleckyCPT,GengCPT,Ni:2007ar,FinelliCPT}.}.
Recently the Wilkinson Microwave Anisotropy Probe (WMAP)
experiment has published the 5-year results for the CMB angular
power spectra which include the TB and EB information
\cite{WMAP51,WMAP52}. They use the polarization power spectra of
WMAP5, TE/TB ($2\leq l \leq450$) and EE/BB/EB ($2\leq l \leq23$),
to determine this rotation angle \cite{WMAPCPT}, and find that
$\Delta\alpha=-1.7\pm2.1$ deg ($1\sigma$).

Besides the WMAP measurement, the BOOMERanG 2003 data also provide
the TB and EB polarization power spectra \cite{B031,B032,B033},
which have been shown to give an interesting constraint on this
rotation angle $\Delta\alpha$ \cite{Feng:2006dp,Xia:2007qs}. Thus
it will be interesting and necessary to combine the full data of
these two experiments for the analysis, which is the aim of this
Letter.


\section{Method and Results}

In our study we make a global analysis on the CMB data with the
public available Markov Chain Monte Carlo package
CosmoMC\footnote{http://cosmologist.info/.} \cite{Lewis:2002ah},
which has been modified to allow the rotation of the power spectra
discussed above, with a new free parameter $\Delta\alpha$. We
assume the purely adiabatic initial conditions and impose the
flatness condition motivated by inflation. In our analysis the
most general parameter space is: ${\bf P} \equiv (\omega_{b},
\omega_{c}, \Theta_{s}, \tau, n_{s}, \log[10^{10}A_{s}], r,
\Delta\alpha)$, where $\omega_{b}\equiv\Omega_{b}h^{2}$ and
$\omega_{c}\equiv\Omega_{c}h^{2}$ are the physical baryon and cold
dark matter densities relative to the critical density,
$\Theta_{s}$ is the ratio of the sound horizon to the angular
diameter distance at decoupling, $\tau$ is the optical depth to
re-ionization, $A_{s}$ and $n_{s}$ characterize the primordial
scalar power spectrum, $r$ is the tensor to scalar ratio of the
primordial spectrum. For the pivot of the primordial spectrum we
set $k_{s0}=0.05$Mpc$^{-1}$. In our calculation we have assumed
that the cosmic rotation angle is constant at all multipoles and
does not depend on $l$. Furthermore, we think that this rotation
angle is not too large and imposed a conservative flat prior
$-\pi/2\leq\Delta\alpha\leq\pi/2$.

In our calculations we combine the full data of WMAP5 and
BOOMERanG 2003 (B03). We calculate the likelihood of CMB power
spectra using the routine for computing the likelihood supplied by
the WMAP\footnote{Legacy Archive for Microwave Background Data
Analysis (LAMBDA), http://lambda.gsfc.nasa.gov/.} and BOOMERanG
groups. Furthermore, we make use of the Hubble Space Telescope
(HST) measurement of the Hubble parameter $H_{0}\equiv
100$h~km~s$^{-1}$~Mpc$^{-1}$ by multiplying a Gaussian likelihood
function $h=0.72\pm0.08$ \cite{Hubble}. We also impose a weak
Gaussian prior on the baryon density
$\Omega_{b}h^{2}=0.022\pm0.002$ ($1\sigma$) from the Big Bang
Nucleosynthesis \cite{BBN}. Simultaneously we will also use a
cosmic age tophat prior as 10 Gyr $< t_0 <$ 20 Gyr.

\vspace{0.5cm}

\begin{figure}
\begin{center}
\includegraphics[scale=0.5]{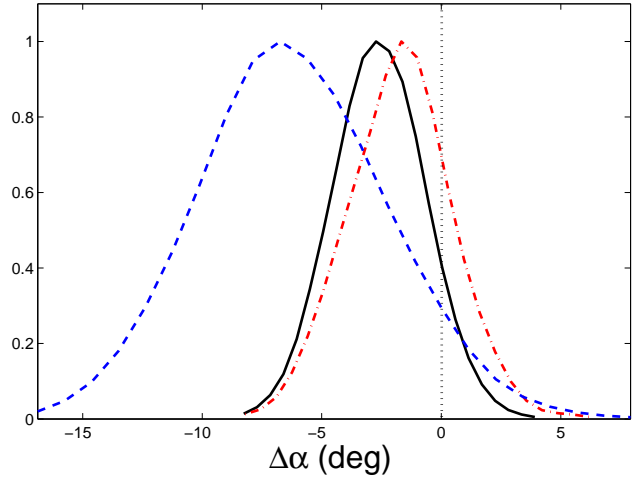}
\caption{One dimensional distributions on the rotation angle
$\Delta\alpha$ from CMB data. The blue dashed line shows our
previous limit on $\Delta\alpha$ from WMAP3 and B03. The red
dash-dot line shows the limit from the full data of WMAP5. The
black solid line is from the full data of WMAP5 and B03. The
horizontal dotted line shows $\Delta\alpha=0$.} \label{Fig}
\end{center}
\end{figure}

Firstly we do a consistency test by comparing two methods used by
us and WMAP group. The WMAP group fixed the parameters except for
$\Delta\alpha$ and $\tau$ in their analysis \cite{WMAPCPT}. The
polarization spectra they considered are TE/TB/EE/BB/EB at $2\leq
l \leq23$ and TE/TB at $24\leq l \leq450$. In our analysis, we
vary all of the parameters in the parameter space and use the full
WMAP5 data including the CMB TT power spectrum. With the WMAP5
data only we find that our result on $\Delta\alpha$ is consistent
with that given by the WMAP group \cite{WMAPCPT}. Therefore, in
the study below, we follow our method to do the calculation with
the combination of the WMAP5 and B03 data.

In Fig.\ref{Fig} we plot our one dimensional constraints on the
rotation angle $\Delta\alpha$ from the CMB data. The blue dashed
line shows our previous result on rotation angle from WMAP3 and
B03 data. The red dash-dot line shows the limit on $\Delta\alpha$
the full data of WMAP5. And the black solid line is our final
result from the full data of WMAP5 and B03 data. The best fit
value of the rotation angle is $\Delta\alpha=-3.5$ deg.
Marginalizing over the posterior distributions of other
parameters, we find that the mean value of the rotation angle is:
\begin{equation}
\Delta\alpha=-2.6\pm1.9~{\rm deg}~(1\sigma)~.
\end{equation}
This constraint is tighter than all of the previous results on
$\Delta\alpha$, say, the error bar is decreased by a factor of 2,
which is profited from the accurate WMAP5 polarization data. On
the other hand, this negative rotation angle is slightly preferred
by the TC and GC information of B03. In the B03 data, the TC power
at $l\sim250$ and $l\sim350$ are both negative, whereas it is
positive at $l\sim450$. The GC power at $l\sim250$, $l\sim350$ and
$l\sim450$ are all negative. Based on the Eq.(\ref{modify}), we
can see that the TC and GC power spectra of B03 really help to
obtain this negative rotation angle.


\section{Summary}

In this Letter we have determined the rotation polarization angle
$\Delta\alpha$ with the combined CMB data from BOOMERanG 2003 and
the newly released WMAP5 data, and obtained
$\Delta\alpha=-2.6\pm1.9$ deg ($1\sigma$), which shows a mild
detection of a nonzero rotation angle and a weak evidence for
cosmological CPT violation. With the near future CMB measurements
our result on the CPT violation could be confirmed or the CPT
symmetry can be verified with a higher precision. For example,
with the
Planck\footnote{http://sci.esa.int/science-e/www/area/index.cfm?fareaid=17/.}
and the Spider measurements \cite{spider} the standard deviation
of the rotation angle will be significantly reduced to
$\sigma=0.057$ deg \cite{Xia:2007gz} and $\sigma=0.38$ deg
\cite{Xia:Spider}, respectively.


\begin{acknowledgements}

We acknowledge the use of the Legacy Archive for Microwave
Background Data Analysis (LAMBDA). Support for LAMBDA is provided
by the NASA Office of Space Science. We have performed our
numerical analysis in the Shanghai Supercomputer Center (SSC). We
thank Yi-Fu Cai, Carlo Contaldi, Eiichiro Komatsu and Tao-Tao Qiu
for discussions. This work is supported in part by National
Natural Science Foundation of China under Grant Nos. 90303004,
10533010 and 10675136 and by the Chinese Academy of Science under
Grant No. KJCX3-SYW-N2. GZ is supported by National Science and
Engineering Research Council of Canada (NSERC).

\end{acknowledgements}


\end{document}